\documentstyle[prl,aps,epsfig]{revtex}
\begin{document}
\twocolumn[\hsize\textwidth\columnwidth\hsize\csname@twocolumnfalse\endcsname
\mbox{}\\]
\noindent
{\bf Comment on "Theory of Unconventional Spin Density Wave:
A Possible Mechanism of the Micromagnetism in U-based Heavy
Fermion Compounds"}

In the recent letter \cite{IkOsh} a new,
very attractive idea is proposed for the explanation of the
micromagnetism in U-based heavy fermion (HF) compounds. For this
sake a nontrivial spin density wave (SDW) state is introduced in
the framework of the Hamiltonian:
\begin{eqnarray}
\nonumber
H=-t\sum_{\langle ij\rangle
\sigma}(c^\dagger_{i\sigma}c_{j\sigma}+h.c.) +
U\sum_in_{i\uparrow}n_{i\downarrow}\\
 -2J\sum_{\langle
ij\rangle}\vec{S}_i\vec{S}_j+(V-\frac{J}{2})\sum_{\langle
ij\rangle \sigma\sigma'}n_{i,\sigma}n_{j,\sigma'}
\label{0}
\end{eqnarray}
where we used the same notations as in \cite{IkOsh} except
$\vec{S}_i=\frac{1}{2}c^\dagger_i \vec{\sigma}c_i$. Unlike the
conventional SDW, the order parameter $\Psi^Q_k\equiv\sum_\sigma
\sigma\langle c^\dagger_{k\sigma}c_{k+Q\sigma}\rangle$ in the
unconventional SDW (d-SDW) \cite{IkOsh} state is characterized by
"d-wave" like k-dependence $\Psi^Q_k\propto\cos k_x - \cos k_y$
\cite{m0}. In this case the ordered staggered magnetic moment
$M_Q$ is equal to zero. The authors restricted themselves to a
very special case of 2D electron system on a simple square
lattice, the shape of the Fermi surface corresponding to the
{\it perfect nesting} with $Q$=$(\pi,\pi)$. The direct and exchange
interaction constants  are chosen positive $V>0$, $J>0$ \cite{m1}.

We are not going to discuss  the origin of the model (\ref{0}) and
criticize its applicability to the essentially 3D
HF compounds (such as $UPt_3$ and $URu_2Si_2$ \cite{GS93}) without
any experimental evidence of perfect or imperfect nesting. 
Our goal is to claim, that even in the
model considered in \cite{IkOsh}, the mean field (MF) analysis
performed by the authors is incomplete and the phase diagram
obtained (see Fig.1 in \cite{IkOsh}) is wrong.

To begin with, let us look carefully on the Hamiltonian (\ref{0}).
One can easily see, that this Hamiltonian  contains the
Coulomb interaction and the {\it ferromagnetic (!)} \cite{m1}
exchange integral. Thus, there are at least four ordered states
which may be realized in this model: itinerant
ferromagnet (FM) state, conventional SDW, charge density wave
(CDW), and d-SDW. One can expect, that the FM state,  missed
by \cite{IkOsh}, will be dominant at least in the limit $U,V \ll J$.
Therefore, to construct a complete phase diagram, the FM state should also
be incorporated into the MF approach.

Let us consider first the case $(U,V,J) \ll t$ when the nesting
property is important and MF analysis is reasonable. The criterion
of instability can be determined from the behavior of the static
response functions \cite{m2}:
$\chi_\alpha(q,0)$=$\chi_\alpha^0(q,0)$/$(1-I_\alpha(q)\chi_\alpha^0(q,0))$,
where $\alpha$=$FM, DW$, $I_{FM}(0)$=$U+4J$, $I_{SDW}(Q)$=$U-4J$,
$I_{CDW}(Q)$=$8V-U-4J$ and $I_{d-SDW}(Q)$=$V$. For the perfect
nesting case $\chi^0_{DW}(Q,0) \sim (1/t) \log^2(t/T)$
\cite{n1}, where one power of logarithm comes from nesting and
another one is due to the Van Hove singularity (VHS).
Nevertheless, $\chi^0_{FM}(0,0) \sim (1/t) \log(t/T)$  is also
singular \cite{n1} due to VHS. The MF critical
temperatures \cite{k1} are  $T^{MF}_{DW}$ $\sim$ $t$
$\exp(-2\pi\lambda_{DW}\sqrt{t/I_{DW}})$ and $T^{MF}_{FM}$ $\sim$ $t$
$\exp( -2\pi\lambda_{FM} t/I_{FM})$, $\lambda_\alpha \sim 1$.
Thus, the FM state certainly wins  when $J\gg (U,V)$ and $V/J$ $\lesssim$ $J/t$ and even overcomes d-SDW in the phase diagram Fig.1 in \cite{IkOsh}. The SDW state is more favorable when $U\gg (V,J)$ and CDW state occurs
when $V\gg (U,J)$. We also emphasize, that unlike VHS, an
additional "nesting" singularity in $\chi(Q, 0)$ is very sensible
to a variety of effects, such as interlayer tunneling, doping, next  hopping, etc, making the application of
model \cite{IkOsh} to real systems nearly impossible.

Let us consider another important limit $U\gg (t,V,J)$, the most
realistic one, since the one-site U should be larger than the
other nearest-neighbor interactions $V, J$ and  $t$ $\sim$
$m_*^{-1}$ $\ll$ $t_0$ ($m_*$ $\gg$ $m_0$ is an effective HF mass,
$m_0$ and $t_0$ correspond to noninteracting fermions). In this
case the $V$ term is irrelevant for the half-filled band due to the
constraint $n_i$=$1$, nesting is not important and only the AF state
with $I_{AF}$ $\sim$ $t^2/U$ \cite{s1} is possible (when $J/t<t/U$).


To conclude, the new d-SDW state predicted in  \cite{IkOsh} cannot be realized for the most physically reasonable limits. The phase diagram in \cite{IkOsh} is wrong, resulting in an
erroneous statement of the d-SDW stability region.
 The very
narrow region of parameters $U$, $V$, $J$, $t$ (which has nothing
to do with those presented in \cite{IkOsh}) where the d-SDW state may
exist requires a more detailed analysis. 

We would like to thank K. Ki\-koin, F. Onu\-fri\-eva
and P. Pfeu\-ty for discussions. M.K. is grateful for the
support and hospitality of Laboratoire L\'eon Brillouin
(CE-Saclay) where this work was carried out.\\
\mbox{}\\
 {\it \noindent
 M.N. Kiselev and F. Bouis\\
 Laboratoire L\'eon Brillouin, CE-Saclay 91191
 Gif-sur-Yvette Cedex, France


\begin{thebibliography}{99}
\bibitem{m0} d-SDW
is more favorable among all other SDW (extended s- and
p- states) due to the Van Hove singularity.
\bibitem{m1} There is not any unconventional SDW state for $J\le 0$.
\bibitem{m2} For simpicity we use the random phase approximation.
\bibitem{k1} There is no true transition to a state with long range order in a 2D system with a continuous symmetry except for $T$=$0$.
\end{thebibliography}
\end{document}